\documentclass{ws-rv9x6}
\usepackage{ws-rv-van}             % numbered citation/references (default)
\newcommand{\be}{\begin{equation}}
\newcommand{\ee}{\end{equation}}
\makeindex
\begin{document}
\def\pmb#1{\setbox0=\hbox{#1}% \kern-.025em\copy0\kern-\wd0
 \kern.05em\copy0\kern-\wd0 \kern-.025em\raise.0433em\box0 }
\chapter{Noncovariant gauges at zero and nonzero temperature\label{ch1}}

\author[P V Landshoff]{P V Landshoff}
%\index[aindx]{Author, F.} % or \aindx{Author, F.}
%\index[aindx]{Author, S.} % or \aindx{Author, S.}

\address{Department of Applied Mathematics and Theoretical Physics\\
Cambridge CB3 0WA\\
pvl@damtp.cam.ac.uk}

\begin{abstract}
I review the formalism for gauge-field perturbation theory in noncovariant
gauges, particularly the temporal axial gauge. I show that, even at zero temperature, there are complications and it is not known whether a formalism exists
for handling these that is correct for all calculations. For thermal field
theory in the imaginary time formalism, there are different difficulties,
whose solution so far is known only up to lowest order in the coupling $g$.
\end{abstract}

\body

\section{Introduction}\label{sec1.1}

Noncovariant gauges in which the gauge field $A_{\mu}$ is defined to satisfy
\be
n.A=n^{\mu}A_{\mu}=0
\ee
where $n^{\mu}$ is some fixed 4-vector of unit length, have been pioneered by Wolfgang Kummer\cite{pioneer}
and later widely used in nonabelian gauge theories\cite{leibbrandt}.
This is partly because Faddeev-Popov ghosts are believed to decouple in such
gauges\cite{kokummer}, and partly because a suitable choice of the direction
of $n^{\mu}$ often seems to simplify calculations. In thermal field theory,
in particular, it is natural to choose, in the rest frame of the ensemble under study,
\be
n^{\mu}=(1,0,0,0)
\ee
The corresponding gauge is called the temporal gauge.

However, there are severe complications with calculating in noncovariant 
gauges. Indeed, it is not even clear that a consistent universal calculation
scheme exists, even for ordinary perturbation theory at zero temperature.
The basic problem is that a naive derivation of the gauge-field propagator
gives
\be
D_{\mu\nu}(k)=\left [-g_{\mu\nu}+{{k_{\mu}n_{\nu}+n_{\mu}k_{\nu}}\over{n.k}}
-n^2{{k_{\mu}k_{\nu}}\over{(n.k)^2}}\right ]{1\over{k^2+i\epsilon}}
\label{propagator}
\ee
and we do not know how to handle the double pole at $n.k=0$. Traditionally,
it was assumed\cite{kummer} that it is correct to apply a principal-value
prescription. This gives the right answer for simple calculations, but
not when there are Feynman graphs where two gauge-field lines carry the
same momentum $k$, so that $D_{\mu\nu}(k)$ has to be squared.  
This was shown\cite{carracc}$^,$\cite{pvlax} by comparing the calculation
in Feynman gauge and in temporal gauge of a Wilson loop in next-to-leading
order, where the sensitive graph is that shown in figure 1.
\begin{figure}
\begin{center}
\epsfxsize = 4.5truecm\epsfbox{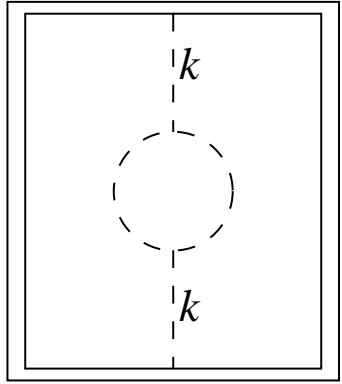}
\caption{Graph in the calculation of a Wilson loop in which a propagator
of momentum $k$ is squared}
\end{center}
\label{wilsonloop}
\end{figure}

Here I will first review progress, or the lack of it, with this difficulty.
I then go on to discuss how things are even more complicated with thermal
field theory in the temporal gauge. 

\section{Temporal gauge at finite temperature}

The propagator (\ref{propagator}) satisfies
\be
n^{\mu}D_{\mu\nu}(k)=0=D_{\mu\nu}(k)n^{\nu}
\label{annihilate}
\ee
In particular, in the temporal gauge only the spacelike components are
nonzero:
\be
D_{ij}(k)={1\over k^2+i\epsilon}\Big (\delta_{ij}-{k_ik_j\over k_0^2} \Big )
\label{temporal}
\ee
The apparent property that 
\be
D_{00}=D_{0i}=D_{i0}=0
\label{simple}
\ee
together with the absence
of ghosts, is why the gauge is supposed to be useful: it makes calculations
much simpler. However, it is not clear that (\ref{simple}) can be assumed to be true.

Certainly, non-leading-order calculations are usually delicate. In my
calculation of the Wilson loop\cite{pvlax}, having established that the
principal-value prescription 
\be
{1\over k_0^2}\longrightarrow {1\over 2}\Big ( {1\over (k_0+i\eta)^2}
+{1\over (k_0-i\eta)^2}\Big )
\label{pv}
\ee
gave the wrong answer, I assumed a prescription that was fairly similar:
\be
{1\over k_0^2}\longrightarrow {1\over k_0^2+\eta ^2}
\label{pseudopv}
\ee
I found that individual graphs diverged as powers of $1/\eta ^2$ 
when $\eta\to 0$. They must be calculated carefully. The limit 
must be taken only right at the end of the calculation, after Feynman's
$\epsilon\to 0$. In particular, powers of $\eta$ from the loop integration
for the graph in figure \ref{wilsonloop} must be retained, as they multiply
powers of $1/\eta$ coming from the $k$ integration. In the end, all
the divergences cancel and the same answer is obtained as in Feynman gauge.

The prescription (\ref{pseudopv}) was chosen so as to give
the correct answer for this calculation, but there is no derivation
of it from first principles and therefore
no guarantee that the same will be true for all other calculations. 

Attempts to derive a prescription typically start with a gauge close
to the one that is wanted. For example\cite{hamiltonian}$^,$\cite{lazzizzera}
one might start by imposing
\def\pd{\partial} \def\pdb{\pmb{{$\partial $}}}
\def\I{{1\over {\pdb}^2}}
\be
A_0 = \eta \, {1 \over{{\pd }_3}} \I\, {\pdb . {\bf A}} =\eta \,{1 \over{{\pd}_3}} \I\, \pdb . {\bf A}^L
\label{constraint}
\ee
with ${\bf A}^L$ the longitudinal field,
$$
A^L_i = \pd _i \I\, {\pdb . {\bf A}}
$$
and let $\eta\to 0$ at the end of any calculation.
One can then eliminate $A_0$ from the Lagrangian, calculate the Hamiltonian
and write down the corresponding equations of motion. Eliminating $A_0$ results
in the loss of an equation of motion, which turns out to be Gauss's
law. When the gauge-field coupling $g$ is switched off, this is just
$\pdb .\dot{\bf A}=0$. This cannot be imposed as an operator
condition. Instead one requires that its matrix element
vanishes: 
\be
\langle P'|\pdb .\dot{\bf A}|P\rangle =0
\label{gauss}
\ee
Here $P$ and $P'$ denote any pair of physical states, and for setting up 
perturbation theory it is sufficient to consider asymptotic states. The equations of motion
ensure that, if this constraint is imposed at any time, say $t=0$, it
remains satisfied at other times.
The constraint (\ref{gauss}) enables one to pick a complete set of
physical states, not uniquely, but most simply one might choose those states
that contain no longitudinal gauge particles. 

Canonical quantisation then leads to the prescription
\be
{1\over k_0^2}\longrightarrow \Big ({1\over {k_0 + i\eta /k_3}} \Big )^2
\label{vienna}
\ee
This is the so-called Vienna prescription\cite{gaigg}. 
It again gives the correct Wilson loop to next-to-leading order\cite{huff},
but again one cannot be sure how generally it may be applied. The problem
is that, until $\eta\to 0$, one cannot be sure that it is valid to
neglect the elements $D_{00},D_{0i},D_{i0}$ of the gauge-field
propagator, nor indeed the ghosts.

The conclusion then is that, attractive as noncovariant gauges may seem
to be, using them for calculations is, to say the least, problematical
even at zero temperature.

\section{Thermal field theory in temporal gauge}

Thermal field theory is formulated by starting with the grand partition
function
\be
Z = \sum_{i} \langle i | e^{-(H - \mu N)/T} |i \rangle
\label{grand}
\ee
from which nearly all (though not all\cite{taylor})
the interesting properties of the system under study
may be calculated.
Here $T$ is the temperature, 
and we use units in which Boltzmann's constant $k_B = 1$.  The
system's Hamiltonian is $H$ and $N$ is some conserved quantum number,
such as baryon number, with $\mu$ the corresponding chemical
potential. The states $|i\rangle$ are a complete orthonormal set of physical states of
the system.  In scalar field theory all states are physical and so
\be
Z = {\rm tr} \; e^{-(H- \mu N)/T}
\ee
which is invariant under changes in the choice of orthonormal basis of
states.  In the case of gauge theories there are unphysical states,
for example longitudinally-polarised photons or gluons, which must be
excluded from the summation in (\ref{grand}).  So then
\be
Z = {\rm tr} \;{\mathbb P}\; e^{-(H-\mu N)/T} 
\label{projection}
\ee
where ${\mathbb P}$ is a projection operator onto physical states.  The
presence of ${\mathbb P}$ can make things more complicated.

For scalar field theory in the so-called imaginary-time formalism of
thermal field theory, the Feynman rules are just as at zero temperature, except
that round each loop of a graph the usual loop-momentum integration
undergoes the replacement
\be
\int {d^4 k\over (2\pi)^4} \rightarrow {iT} {\sum_n} \int
{d^{3}k\over (2\pi)^3} 
\label{expansion}
\ee
Here the summation is over discrete values 
\be
k_0=n\pi T~~~~~~ n=0,\pm 2,\pm 4,\dots
\label{discrete}
\ee
which has the consequence that the propagator is periodic in the
time $t$:
\def\k{{\bf k}}
\be
D(t,\k )=D(t-i/T,\k )
\label{periodic}
\ee

Were this rule to apply also to a gauge-field theory in the temporal
gauge, there would obviously
be a difficulty, since the summation would include a contribution from
$n=0$, that is $k_0=0$, where the zero-temperature propagator has a double pole.

The transverse gauge-field
propagator does behave similarly to a scalar field, with
\be
[D_n^T(\k )]_{ij}=-\left (\delta _{ij} - {{k_ik_j}\over{\k ^2}}\right )
   {1\over {\pi ^2 n^2 T^2 + \k ^2}}  ~~~~~~ n=0,\pm 2,\pm 4,\dots
\label{transverse}
\ee
and it has the
periodicity (\ref{periodic}).  
However\cite{james}, this is not the 
case for the   
the longitudinal field: because of the presence of the projection
operator in (\ref{projection}), canonical quantisation results in a
longitudinal propagator $[D^L(t,\k )]_{ij}$ that is not periodic in $t$. 
But there is a simplification to compensate for this complication:
$[D_n^L(\k )]_{ij}$ does not have a double pole at $n=0$. Indeed, it
is regular there, but $n$ is not restricted to even values:
$[D_n^L(\k )]_{ij}= (k_ik_j/\k ^2) D_n^L(k_3 )$, with
\be
D_n^L(k_3)=\left\{\begin{matrix}{1/({4T^2)}}&n=0\\ 
{-i{\epsilon(k_3)}/({2\pi nT^2})}&n{\hbox{ even}}\\
{-1/({\pi ^2n^2T^2})}+{i{\epsilon(k_3)}/( {2\pi nT^2})}&n{\hbox{ odd}}
\end{matrix}\right . 
\ee
The non-vanishing of $[D_n^L(\k )]_{ij}$ for odd $n$ means that it is easier
not to work with $[D_n(\k )]_{ij}$ and perform summations over the various
$n$ associated with the different lines in a Feynman graph, but instead
to work with propagators $[D^L(t,\k )]_{ij}$ and integrate over the times
associated with the various vertices. 

Another complication that has to be taken into account is the constraint 
(\ref{gauss}) on the physical states. At zero temperature, or when one is 
working in the real-time thermal formalism, one uses asymptotic states and
this  constraint is sufficient.
But the imaginary-time formalism rather uses interaction-picture states
and  so the Gauss operator is now
\be
G^a(t, {\bf x})= \pdb .\dot{\bf A}^a(t, {\bf x}) -gf^{abc} {\bf A}^b(t, {\bf x}).\dot {\bf A}^c(t, {\bf x})
\label{gaussop}
\ee
and it turns out\cite{james} that one needs to impose a set of 
constraints
\be
\langle P'|\prod _{i=1}^N G^{a_i}(0,{\bf x})|P\rangle=0~~~N=1,2,\dots
\label{gausser}
\ee
As one increases the accuracy of one's calculation to higher powers of $g$,
one needs to go up to higher and higher values of $N$. The physical states
can no longer be taken as those with purely-transverse gauge particles.
The solution of (\ref{gausser}) is known only for low-order calculations
for which it is sufficient to go up to $N=2$.

\bibliographystyle{ws-rv-van}
\bibliography{ws-rv-sample}

\begin{thebibliography}{9}        % for non BIBTeX users 
\bibitem{pioneer} W Kummer, Acta Physica Austriaca, 14 (1961) 149
\bibitem{leibbrandt} G Leibbrandt, Reviews of Modern Physics 59 (1987) 1067
\bibitem{kokummer} W Konetschny and W Kummer, Nuclear Physics B100 (1975) 106 and B124 (1977) 145 
\bibitem{kummer} W Kummer, Acta Physica Austriaca 41 (1975) 315
\bibitem{carracc} S Carracciolo, G Curci and P Menotti, Physics Letters
113B (1982) 311
\bibitem{pvlax} P V Landshoff, Physics Letters B169 (1986) 69
\bibitem{hamiltonian} P V Landshoff, Physics Letters B227 (1989) 427
\bibitem{lazzizzera} I Lazzizzera, Physics Letters B210 (1988) 188
\bibitem{gaigg} P Gaigg and M Kreuzer, Phys Lett B205 (1988) 530
\bibitem{huff} H H\"uffel, P V Landshoff and J C Taylor, Physics Letters B217 (1989) 147
\bibitem{taylor} P V Landshoff and J C Taylor, Nuclear Physics B430 (1994) 683
\bibitem{james} K A James and P V Landshoff, Physics Letters B251 (1990) 167

\end{thebibliography}
\printindex

\end{document}